\documentclass[%
 twocolumn,
 amsmath,amssymb,
 aps,
]{revtex4-2}

\usepackage{graphicx}%
\usepackage{dcolumn}%
\usepackage{bm}%
\usepackage[parfill]{parskip}
\usepackage{epstopdf}
\usepackage{float}
\usepackage{tikz}
\usetikzlibrary{arrows}
\usepackage{xcolor}
\usepackage{braket}
\usepackage{subcaption}
\usepackage{graphicx}
\usepackage{ragged2e}
\usepackage{caption}

\definecolor{dred}{HTML}{AE2D2D}

\begin{document}

\title{Physical coherent cancellation of optical addressing crosstalk in a trapped-ion experiment}%
\author{Jeremy Flannery}%
\author{Roland Matt}
\author{Luca Huber}
\author{Kaizhao Wang}
\author{Christopher Axline}
\author{Robin Oswald}
\author{Jonathan P. Home}%
\affiliation{
Institute for Quantum Electronics, ETH Zurich, Otto-Stern-Weg 1, 8093 Zurich, Switzerland}

\date{\today}%

\begin{abstract}
We present an experimental investigation of coherent crosstalk cancellation methods for light delivered to a linear ion chain cryogenic quantum register. The ions are individually addressed using focused laser beams oriented perpendicular to the crystal axis, which are created by imaging each output of a multi-core photonic-crystal fibre waveguide array onto a single ion. The measured nearest-neighbor native crosstalk intensity of this device for ions spaced by 5 $\mu$m is found to be $\sim 10^{-2}$. We show that we can suppress this intensity crosstalk from waveguide channel coupling and optical diffraction effects by a factor $>10^3$ using cancellation light supplied to neighboring channels which destructively interferes with the crosstalk. We measure a rotation error per gate on the order of $\epsilon_{x} \sim 10^{-5}$ on spectator qubits, demonstrating a suppression of crosstalk error by a factor of $> 10^2$. We compare the performance to composite pulse methods for crosstalk cancellation, and describe the appropriate calibration methods and procedures to mitigate phase drifts between these different optical paths, including accounting for problems arising due to pulsing of optical modulators. 
\end{abstract}

\maketitle

\section{INTRODUCTION}\label{Sec:Intro}

The progression of quantum information processing using trapped ions has seen single and two-qubit gate fidelities surpass quantum error correcting thresholds \cite{Ballance2016, Gaebler2016, Blume2017, Srinivas2021, Clark2021}. However, in integrating this control into larger systems, other types of error become apparent.  One primary error of this type is crosstalk \cite{Huang2020, Debroy2020, Sarovar2020}, in which control fields for a target qubit leak to nearby spectator qubits due to imperfections \cite{Sarovar2020}. Not only can crosstalk lead to significant errors but the errors are correlated. This imposes challenges for the implementation of fault-tolerant quantum computation where a key design principle is that errors should be and remain localized \cite{Parrado2021}. Various approaches to suppress errors due to crosstalk have been developed, such as composite pulses \cite{Merrill2014, Nigg2014, Fang2022}, dynamical decoupling \cite{Fang2022} and optimal control \cite{Winick2021, Carvalho2021}. These methods are algorithmic in nature: they live with a certain level of physical crosstalk and utilize control methods to suppress the logical effect. 

For quantum computing systems based on a linear chain of trapped-ion qubits in a linear chain, individual control is performed by applying tightly focused beams to specific ions \cite{Binai2023, Sotirova2023}. Here we consider a popular approach where the radial modes of the ion crystal are used to generate entanglement. This is accomplished by orienting the addressing beams along the direction perpendicular to the ion chain. The beam spot size is fundamentally limited by diffraction, which naturally causes crosstalk to arise by the undesired partial illumination of neighbouring ions. The ion spacing within the chain will thus dictate the amount of crosstalk to neighboring qubits, with lower values typical as inter-ion spacing is increased. This competes with the requirements for fast gates however, which favour close spacing of the ions. While we consider below only linear-chain approaches to trapped-ion quantum computing, we note that separating ions by shuttling can also reduce crosstalk. Nevertheless crosstalk has still been observed in such systems with both free space \cite{Pino2021} and integrated optics \cite{Mehta2020, Mordini2024}.

In this work we investigate suppression of optical crosstalk by physical coherent cancellation (PCC). In this approach an out-of-phase optical ``cancellation'' pulse is applied to a spectator site in parallel to the application of a gate pulse on the target ion, with the aim of causing destructive interference between the crosstalk and the cancellation light. This requires interferometric stability between the addressing and cancellation light. PCC is a versatile technique that can be applied to other quantum computing platforms that possess phase-coherent crosstalk \cite{Mundada2019, Nuerbolati2022}. A similar approach was previously used in the context of microwave gates on trapped ions \cite{Aude2017}.

We begin in Section \ref{Sec: Exp setup} with a description of the experimental apparatus, focusing on the single-ion addressing system utilised to deliver both the gate pulses as well as the crosstalk compensation pulses. The theoretical description of PCC is presented in Section \ref{Sec: Theory}, providing a framework to determine the necessary calibration requirements for reducing crosstalk errors below a given threshold. Section \ref{Sec: noise} covers the performance of this technique, in particular slow optical phase drifts from temperature and pressure fluctuations, as well as fast duty-cycle related effects. We also present methods used to reduce the impact of these noise effects on the cancellation fidelity. Finally, in Section \ref{Sec: results} the crosstalk errors of our device are characterized, and the performance of PCC in our setup is compared to composite pulse crosstalk suppression techniques.

\section{Experimental Setup} \label{Sec: Exp setup}

\begin{figure}[ht]
    \centering
        \includegraphics[scale=0.2]{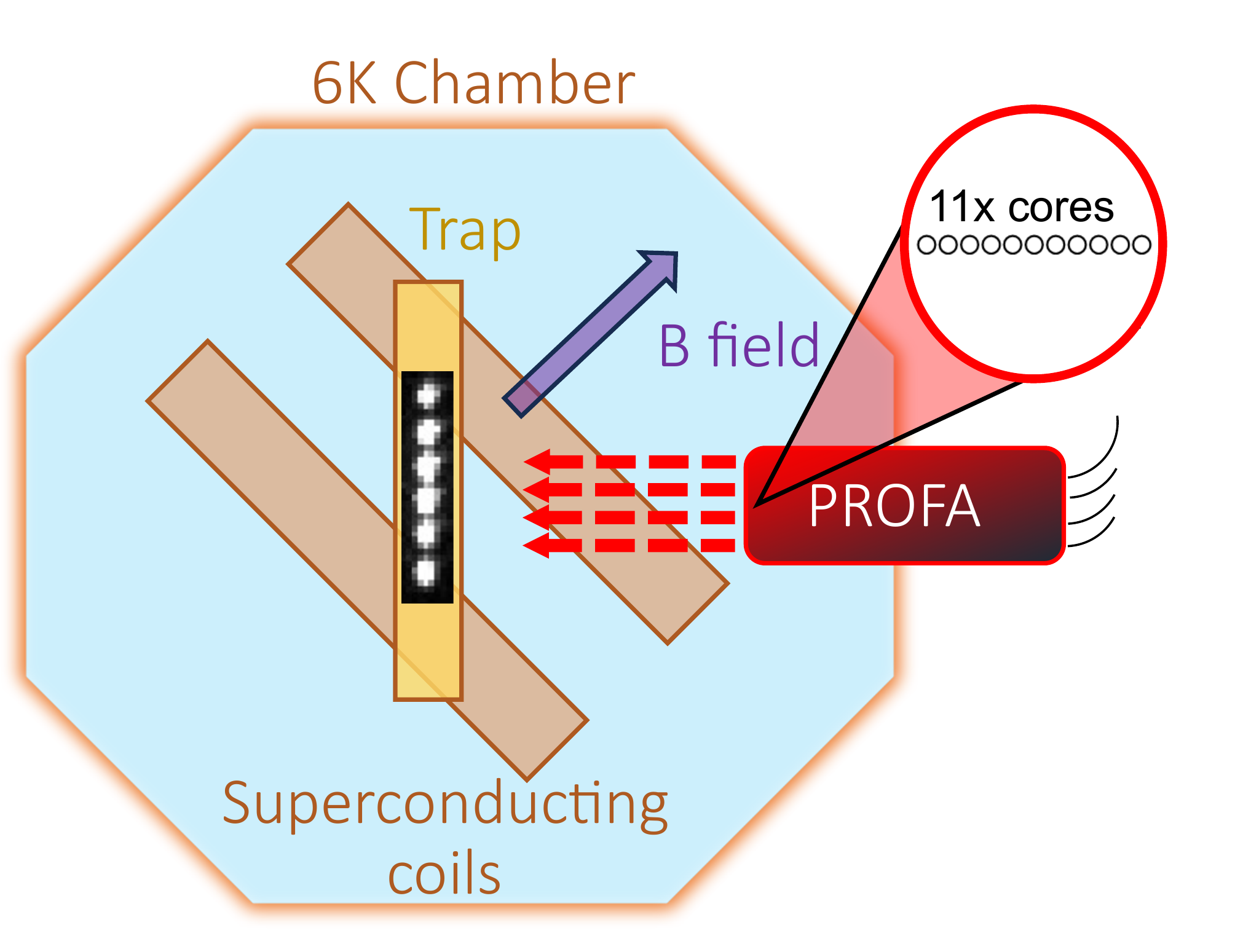} 
    \caption{Illustration of the experimental setup. The single-ion addressing PROFA fiber is oriented perpendicular to the axis of the ion chain trapped using a monolithic ion-trap. This multi-core fiber has 11 individually controlled cores and is mounted directly to the cryogenic chamber. A passively stable magnetic field is provide by NdTi superconducting coils \cite{Oswald2022}.}
    \label{fig:exp setup}
\end{figure}

\begin{figure}
    \centering
        \includegraphics[scale = 0.2]{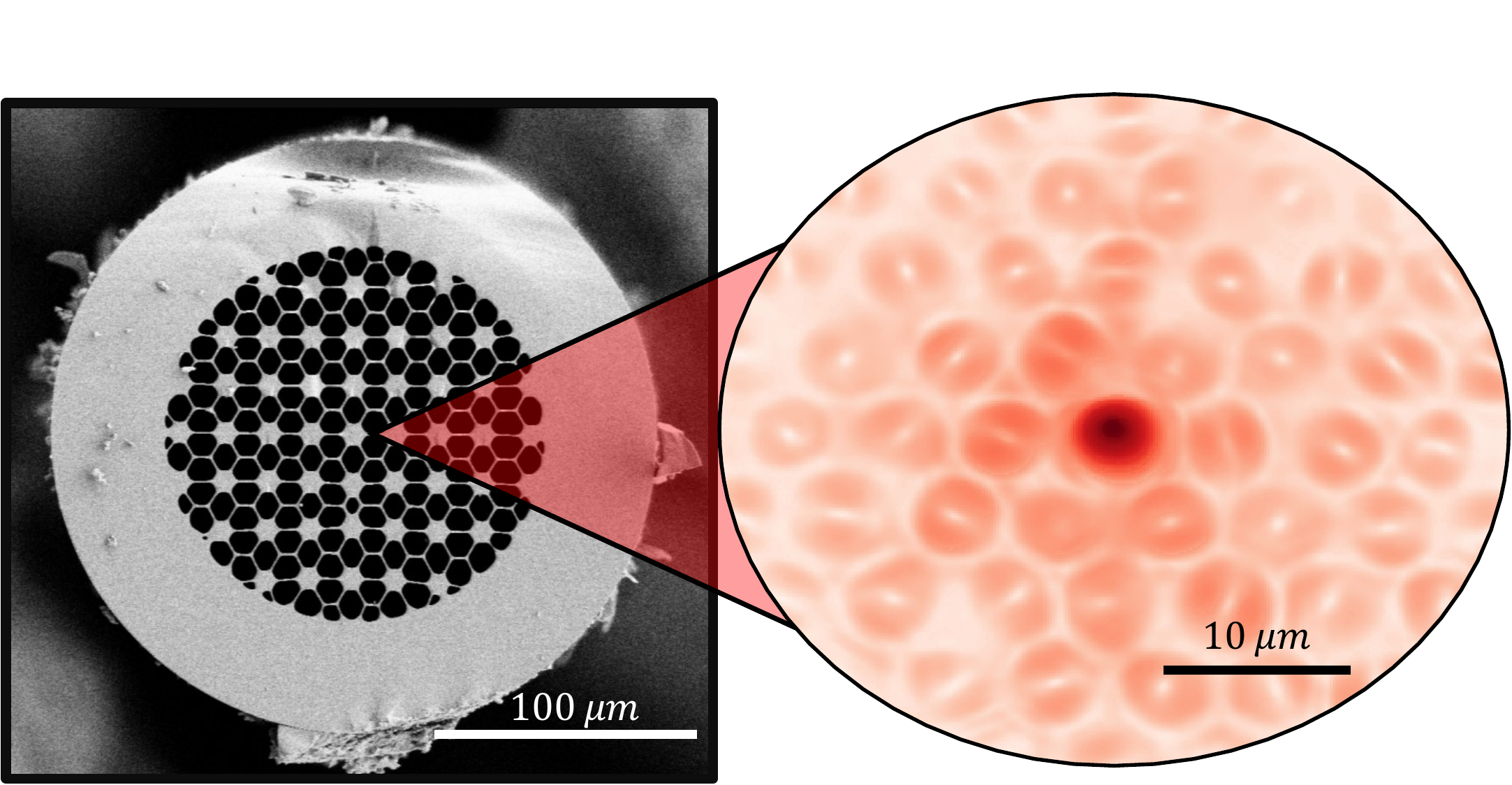}
    \caption{(a) Scanning electron microscope (SEM) image of the cross-sectional profile of the Pitch Reducing Optical Fiber Array (PROFA), which is a type of photonic crystal fiber designed with a pitch of 5 $\mu$m between neighbouring addressing cores, with mode sizes of $\sim 2 \mu$m. (b) The intensity profile of the PROFA device with one out of the 11 individually controllable cores illuminated with light. The neighbouring cores experience crosstalk illumination due to evanescent cross-coupling between cores, internal scattering, and excitation of higher order spurious modes.}
    \label{fig: profa}
\end{figure}

The experimental platform used for this work is a cryogenic trapped-ion setup \cite{Decaroli2021, Oswald2022} sketched in Fig. \ref{fig:exp setup}. The ion addressing system is realized using a Pitch Reducing Optical Fiber Array (PROFA) manufactured by Chiral Photonics, which is placed within the cryogenic vacuum system \cite{Matt2023}. The output end of this device, shown in  Fig. \ref{fig: profa}, is a multi-core photonic crystal fiber where each core guides light fed independently from an input optical fibre. The light emitted from the end of the multi-core PCF is imaged with a 1:1 magnification telescope arrangement onto the ion string. Using independent fibre AOMs with polarization tuning, we control light feeding each core separately,  allowing addressing at each ion location with polarization, frequency, phase and amplitude control.
The 11 individually controllable cores are arranged linearly with a  5 $\mu$m pitch and an approximately 1.6 $\mu$m mode spot size. Since the multi-core fiber is mounted within the vacuum chamber,  vibrations in the experiment chamber are common mode between cores, and also common mode with respect to the distance between the output of the muti-core PCF and the ion trap. This makes the system a robust way of delivering light to ion strings in a cryogenic environment. The improved stability of the waveguide addressing compared to free space optics is demonstrated by a considerable enhancement of the observed optical qubit coherence by Ramsey coherence measurements, increasing from 3 ms to 70 ms.

\begin{figure}
    \centering
         \begin{subfigure}[b]{0.5\textwidth}
             \centering
            \caption{}
            \includegraphics[scale=0.14]{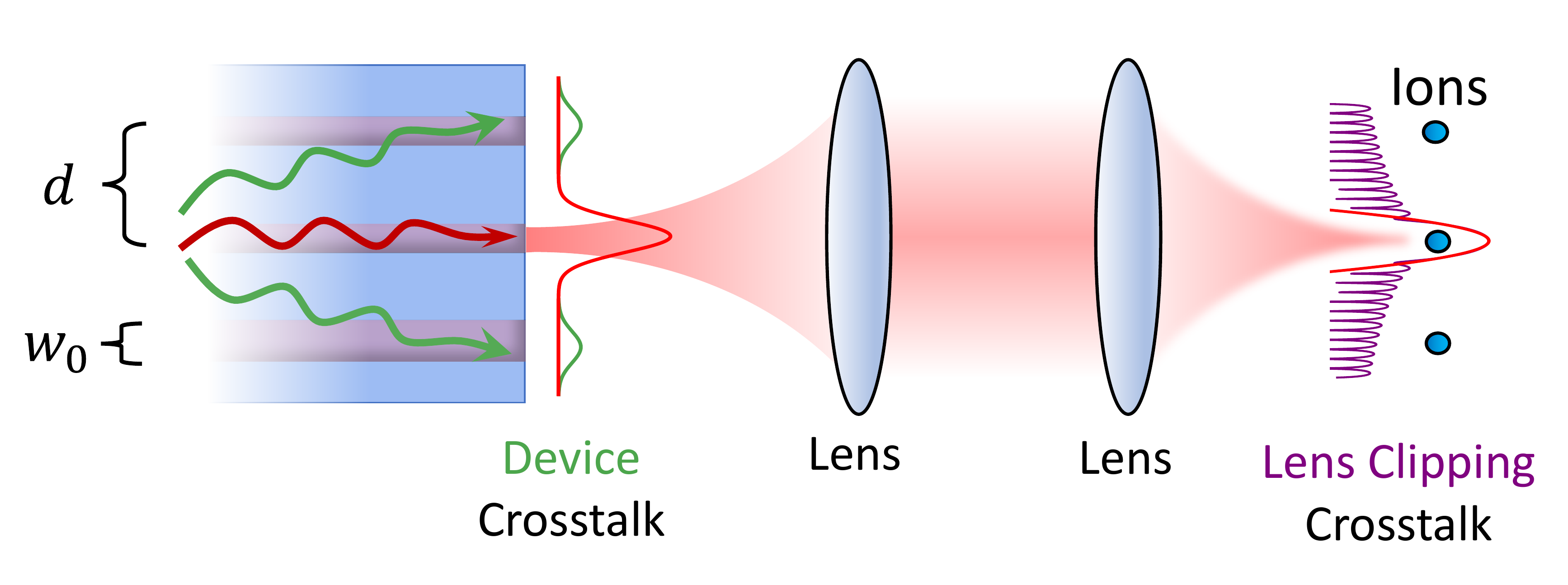}
             \label{fig: crosstalk type}
         \end{subfigure}
    \hfill
         \begin{subfigure}[b]{0.4\textwidth}
             \centering
            \caption{}
             \includegraphics[scale = 0.4]{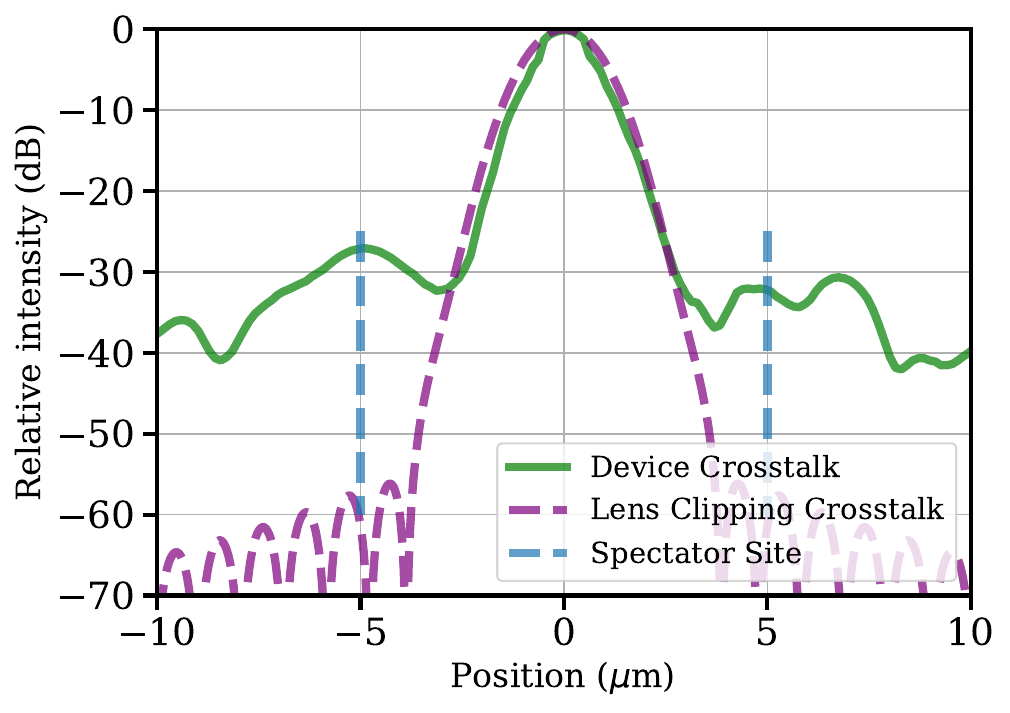}
             \label{fig: intensity crosstalk}
         \end{subfigure}
    \caption{(a) A depiction of the two relevant causes of crosstalk present in the addressing system. (1) Device crosstalk: caused by evanescent coupling of addressing light into the neighbouring cores of the addressing waveguide device and (2) Lens clipping crosstalk: due to the effect of a finite lens aperture size, which deforms the focused spot profile from a perfect Gaussian (red curve) to one in which a much larger intensity in present in the tails of the focused spot (purple curve). (b) Intensity crosstalk produced by the PROFA device that is used to create the individually addressing tightly focused spots. The device crosstalk (green solid line) was measured and found to be the limiting factor compared to the lens clipping crosstalk (purple dashed line), which was analytically estimated (for a lens NA = 0.35, and a spot size of 1.6 $\mu$m at 729 nm wavelength). At a spectator ion separation of approximately 5 $\mu$m, an intensity crosstalk was to measured to be $> 10^{-3}$.}
\end{figure}

 Two main types of crosstalk occur in this approach. The first is caused by  cross-coupling between the cores within the multi-core PCF device (it is not clear to us where this occurs) - we will refer to this in what follows as device crosstalk. The effect of this can be seen from a profile of the intensity output when one core is illuminated with light, which is shown in the inset of Fig. \ref{fig: profa}. The second source of cross talk is due to diffractive effects from lens clipping due to the finite numerical aperture of $\sim0.38$ used for the subsequent free-space relay optics used to focus the beams on the ion - we refer to this as diffractive crosstalk. This diffraction pattern can be calculated analytically by the use of the Kirchoff-Huygens formula \cite{Tanaka1985}. The characteristic attribute of this lens clipping effect is that the central peak width becomes broadened, and the intensity envelope of the sidelobes decay with distance much weaker than an ideal Gaussian profile (purple plot in Fig. \ref{fig: intensity crosstalk}). 

Both the device crosstalk as well as the diffractive crosstalk need to be evaluated for any general single-ion addressing system. In the PROFA, the device crosstalk dominates. Fig. \ref{fig: intensity crosstalk} shows the measured device crosstalk (green line), performed by imaging the tip of the multi-core fiber with a high NA objective - intensity crosstalk at the $\sim 10^{-3}$ level is observed for ions spaced by 5 micron. The analytically calculated lens clipping (purple line) is relatively weak by comparison, in part due to the need for only a 1-to-1 magnification. It is possible to attempt to decrease device crosstalk by increasing  the core spacing of the addressing device. In considering upgraded devices of this type, we found a trade-off due to the increased demagnification, which tends to increase the diffraction crosstalk.

\section{Crosstalk Cancellation} \label{Sec: Theory}

We now consider cancellation of the crosstalk light at the location of the spectator ion by destructive interferance with light applied to the spectator addressing core. Exact cancellation requires  matching of both the amplitude and phase of the crosstalk, and that the polarizations of the two light fields at the ion are the same. We next examine how imperfections of amplitude and phase affect the residual crosstalk. We define the complex-valued Rabi frequency experienced by the spectator ion due to crosstalk as $\Omega_{\rm CT}$.
For a compensation tone producing a Rabi frequency of $\Omega_{\rm comp}$, the effective crosstalk Rabi frequency, $\Omega_{\rm eff}$, applied to the spectator ion is the complex-value sum of both the crosstalk and compensation electric fields,
\begin{equation} \label{eq: Rabi eff}
    \Omega_{\rm eff} = \Omega_{\rm CT} + \Omega_{\rm comp} \ .
\end{equation}
Defining the relative strength, $f_{\rm comp}$, of the compensation field amplitude as $|\Omega_{\rm comp}| = f_{\rm comp}|\Omega_{\rm CT}|$, and the relative phase $\Delta \phi$ between the crosstalk and compensation fields, we can rewrite Eq. \ref{eq: Rabi eff} as 
\begin{align*}
    \Omega_{\rm eff}  = |\Omega_{\rm CT}|e^{i\omega_Lt} (1 + f_{\rm comp}e^{i \Delta \phi})   
\end{align*}
with magnitude 
\begin{equation} \label{eq: CT compensation amp}
    |\Omega_{\rm eff}| = |\Omega_{\rm CT}|\left|1 + f_{\rm comp}e^{i \Delta \phi}\right| \ ,
\end{equation}
producing zero for $f_{\rm comp} = 1, \Delta \phi = \pi$. The amplitude of the crosstalk is reduced as long as
\begin{equation} \label{eq: crosstalk inequality}
    \left|1 + f_{\rm comp}e^{i \Delta \phi}\right| = \sqrt{1 + f_{\rm comp}^2 + 2f_{\rm comp} \cos(\Delta\phi}) < 1
\end{equation}
which places the requirement on the phase that 
\begin{equation}
    \cos(\Delta \phi) < \left(-\frac{f_{\rm comp}}{2}\right).
\end{equation}
For $f_{\rm comp} = 1$, the break-even point of the compensation phase is then $2\pi/3 < \Delta \phi < 4\pi/3$ (a range of $\pm\pi/3$ around the optimal value of $\Delta\phi=\pi$).

Since the target ion is illuminated by significant light intensity, it also experiences AC Stark shifts, but for the spectator ion these are close to zero. It is therefore common to detune the gate laser pulses to the Stark-shifted resonance, meaning that the crosstalk light will have a detuning of $\Delta_{\rm CT}$ from the spectator ion resonance.

Under these conditions the spectator ion is rotated by a unitary $\hat{R}_{\rm CT}$ away from its initial state (neglecting any dephasing of the crosstalk light). The overlap of the rotated spectator qubit with its initial state, $|\psi_0\rangle$ is given by
\begin{equation}
    |\langle\psi_0|\hat{R}_{\rm CT}|\psi_0\rangle|^2= 1 - \epsilon
\end{equation}
where we define $\epsilon$ as the rotation error. If the initial state Bloch vector is orthogonal to the axis of rotation, the rotation error is maximized such that
\begin{equation} \label{eq: error}
    \epsilon_{\rm max} = \sin^{2}\frac{|\tilde{\Omega}_{\rm eff}|}{2}t
\end{equation}
with $\tilde{\Omega}_{\rm eff}^2 \equiv {\Omega}_{\rm eff}^2 + \Delta_{\rm CT}^2$ defined as the generalized effective Rabi frequency.

The average fidelity, $F_{\rm avg}$,  can be found by averaging over all initial states with equal weighting \cite{Mayer2018}, giving an error $1 - F_{\rm avg}$ that is less than the above $\epsilon_{\rm max}$. This can be easily realized by the fact that initial states that are close to eigenstates of $\hat{R}_{\rm CT}$ act to increase the average fidelity. To characterize the behaviour of rotation errors we will continue our analysis assuming the error given by Eq. \ref{eq: error}.

\begin{figure}
\centering
    \begin{subfigure}[b]{0.4\textwidth}
         \centering
            \caption{} \label{fig: compensation phase 1d plot}
            \includegraphics[scale=0.4]{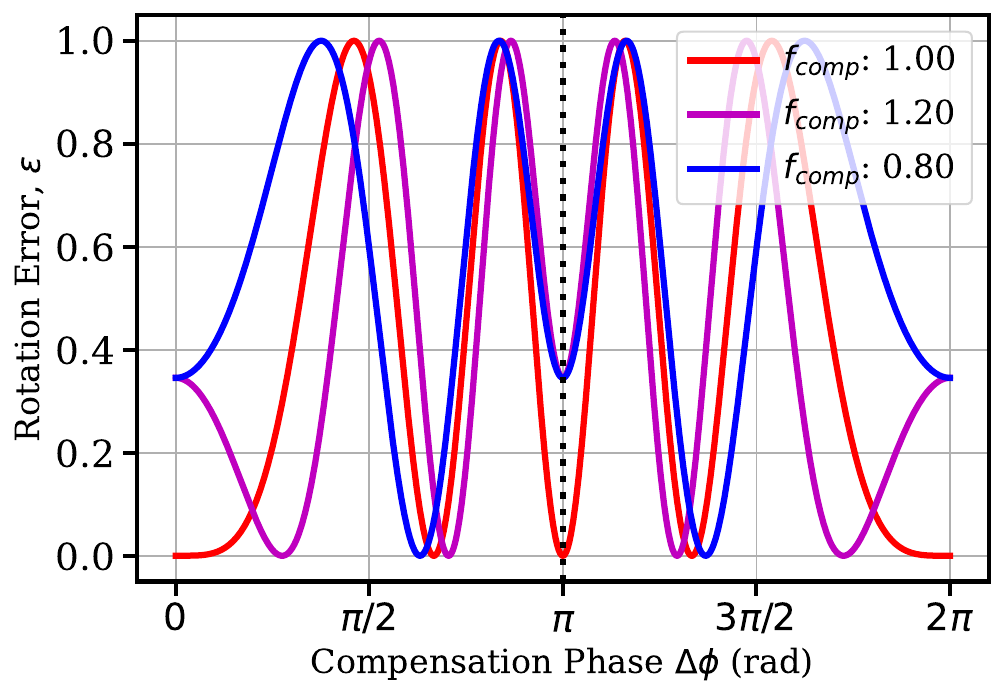}
    \end{subfigure}
    \hfill
    \begin{subfigure}[b]{0.4\textwidth}
            \centering
        \caption{} \label{fig: phase calibration}
        \includegraphics[scale=0.3]{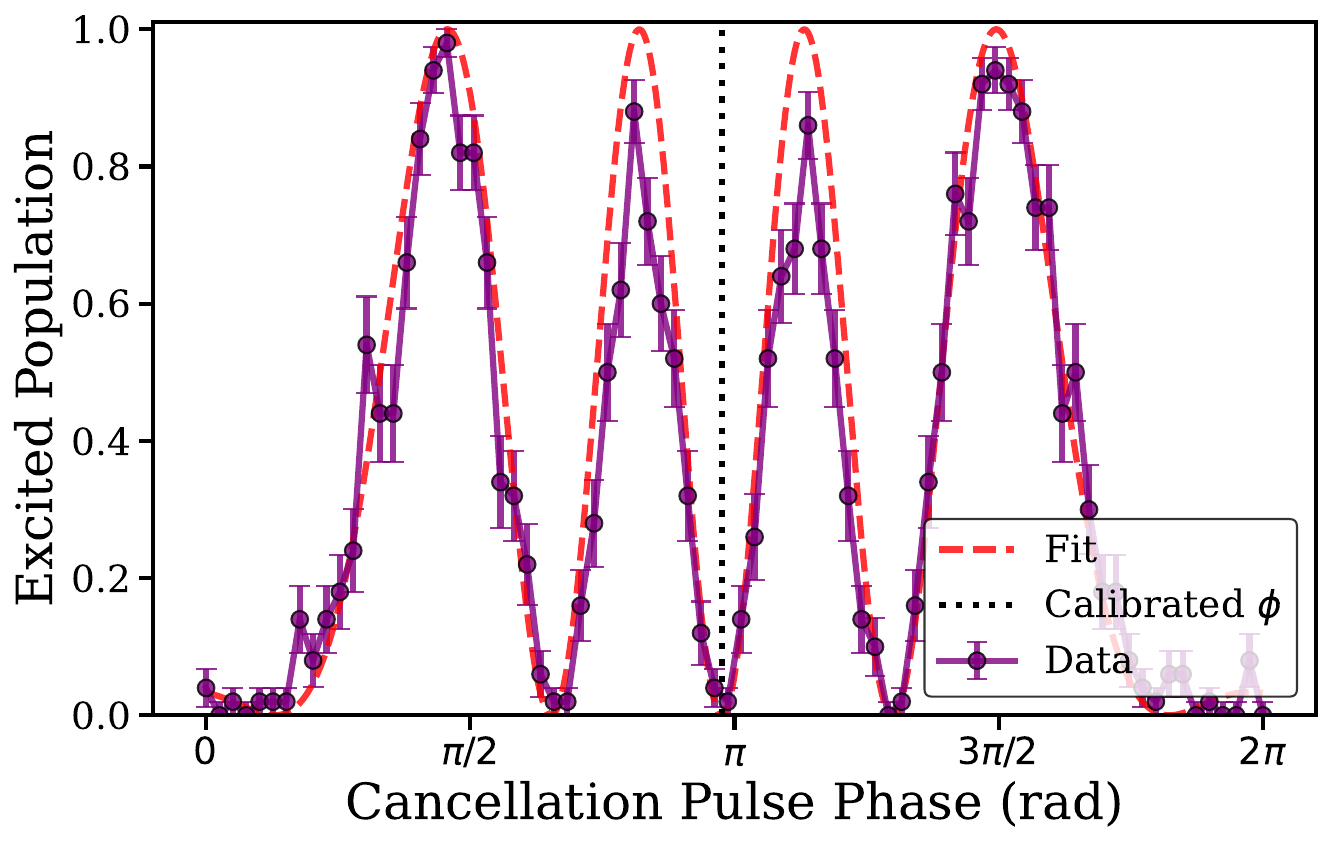}
     \end{subfigure}
    \caption{
    Population of the spectator ion as a function of $f_{\rm comp}, \Delta \phi$ at a pulse duration of twice the crosstalk $\pi$ time, $t = 2t^{CT}_{\pi}$, as used for calibration of the optimal phase. (a) Theoretical curve. The dotted black line represents the ideal crosstalk compensation phase, $\Delta\phi =\pi$. (b) Phase calibration of a PCC pulse. The data is fitted using Eq. \ref{eq: error} (fit result given as dashed-red curve) to determine the optimal compensation phase required to extinguish crosstalk intensity, shown as the black dotted line.}

\end{figure}

Fig. \ref{fig: compensation phase 1d plot} shows the error at different ratios of the compensation pulse to crosstalk light amplitudes $f_{\rm comp}$ as a function of $\Delta \phi$, which we use to calibrate the phase. To amplify the error we choose a particularly long pulse time of $t = 2 t^{CT}_{\pi}=2 \frac{\pi}{\Omega_{CT}}$. The black dotted line shows the correct compensation phase. At $\Delta\phi=0$ the compensation light constructively interferes with the crosstalk light, thus for $f_{\rm comp} = 1$ the ion undergoes a $4\pi$ Bloch sphere rotation. The other point of interest is the minima around $\Delta \phi = \pi$, which we use to set the value of the phase. The accuracy of the phase calibration can be improved by simply setting the pulse time to be many $\pi$ times ($t=2nt_{\pi}, n\in\mathbb{N}$) narrowing the feature at $\Delta\phi = \pi$.

\begin{figure}
\centering
        \includegraphics[scale=0.3]{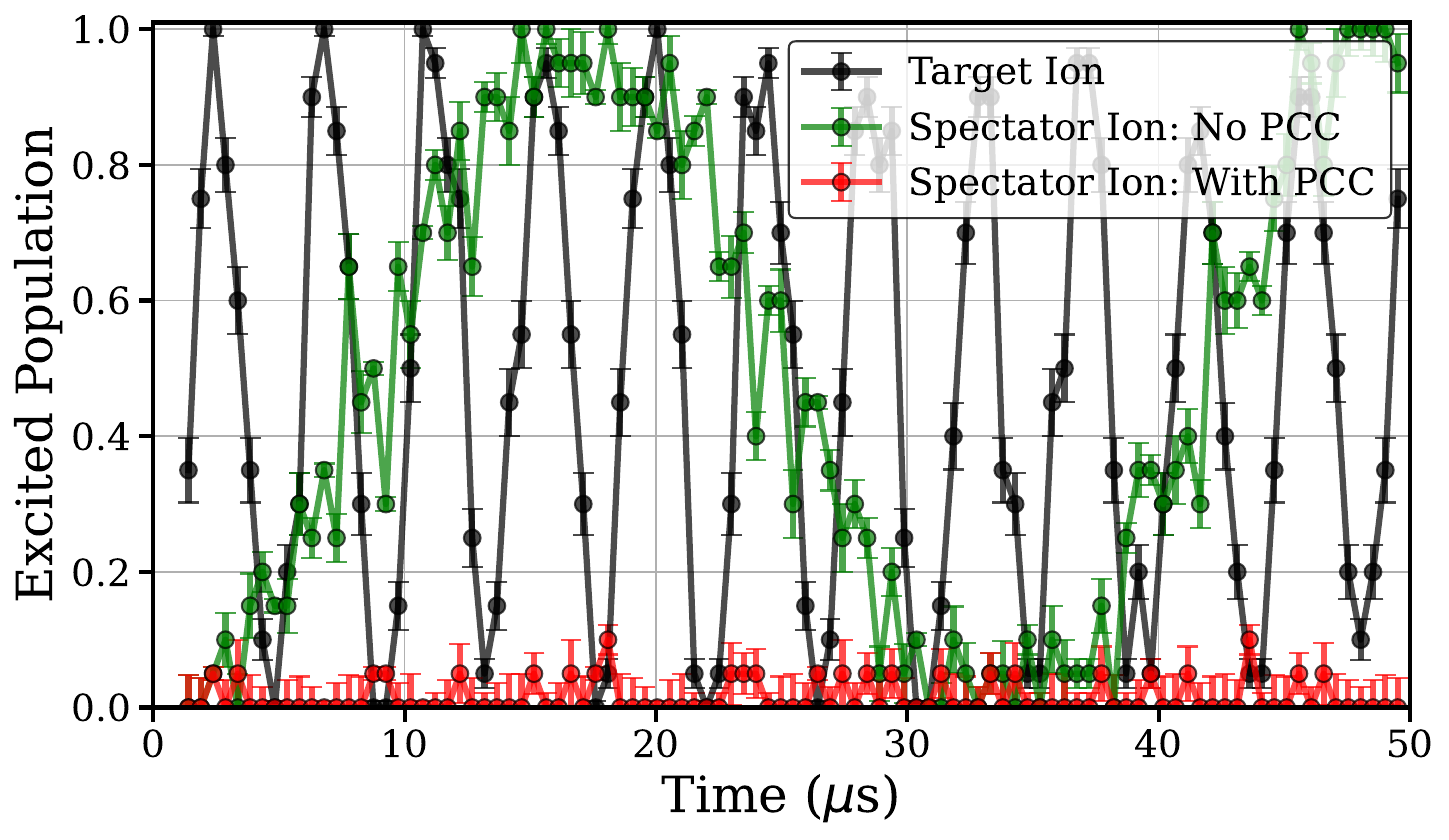}     
\caption{Rabi flopping of the target ion (blue) and spectator ion (green) while only the target addressing beam is applied. The crosstalk can then be cancelled by simultaneously applying the calibrated compensation pulse on the spectator ion (red) to eliminated the light intensity at the spectator location.}
\label{fig: Rabi cancellation}
\end{figure}

Experimental data is shown in Fig. \ref{fig: phase calibration} . The calibrated phase that extinguishes the crosstalk light is given by the black dotted line. The effect of applying this physical cancellation pulse is evident in Fig. \ref{fig: Rabi cancellation}. The black and green data show Rabi flopping of the target and spectator ion excited state populations, respectively, when only the target gate pulse is applied. When the PCC tone is applied to the spectator ion, the Rabi rate of the effective crosstalk is reduced by a factor of $\sim 40$.

\begin{figure}
\centering
    \includegraphics[scale=0.4]{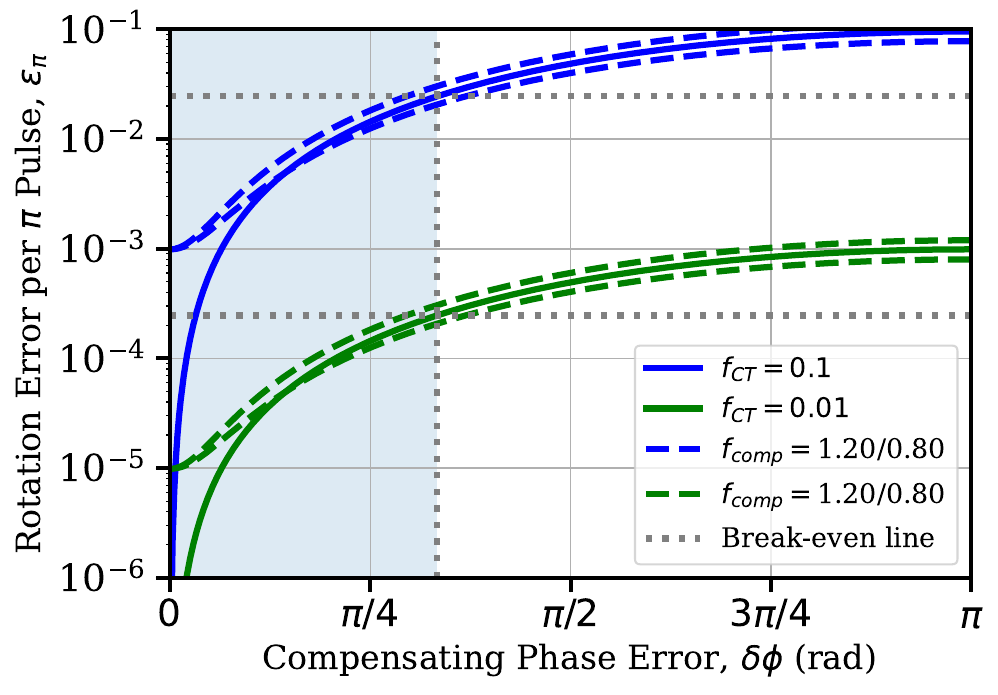}
    \caption{Rotation error after a single target $\pi$ pulse, $t=\pi/\Omega_0$, as described by Eq. \ref{eq: pi pulse error}, for a varying error in the compensation pulse phase from the optimal value, $\delta \phi = |\pi - \Delta \phi|$. The blue and green lines show the crosstalk error on the spectator ion for a relative crosstalk amplitude, $f_{\rm CT}=\Omega_{\rm CT}/\Omega_0$, of $10\%$ and $1\%$, respectively. The solid lines represent a perfect compensation amplitude matching, $f_{\rm comp} = 1$, while the dashed lines show when the compensation amplitude is miscalibrated by $\pm20\%$. For a perfectly matched compensation amplitude, the error in phase required to reduce the crosstalk below the uncompensated crosstalk error is $\delta\phi<\pi/3$ (light blue).}
    \label{fig: compensation error}
\end{figure}

We now investigate the general behaviour of the rotation errors described by Eq. \ref{eq: error}. For simplicity, we set the spectator crosstalk detuning $\Delta_{\rm CT} = 0$. The crosstalk error for a time $Nt_{\pi}$ on the spectator qubit, $\epsilon_{N\pi}$, can then be found by setting $t = N t_{\pi}$, to give 
\begin{equation} \label{eq: pi pulse error}
    \epsilon_{N\pi}(N, f_{\rm eff}) = \sin^{2}\left(\frac{\pi}{2}N f_{\rm eff}\right)
\end{equation}
where $
f_{\rm eff} = f_{\rm CT} \sqrt{1 + f_{\rm comp}^2 + 2 f_{\rm comp} \cos(\Delta\phi)}$ for a relative crosstalk ratio of $f_{CT} =\Omega_{\rm CT}/\Omega_0$, with the Rabi frequency, $\Omega_0$, applied to the target ion. Fig. \ref{fig: compensation error} depicts the error for a $\pi$ pulse ($N=1$) from Eq. \ref{eq: pi pulse error} with $10\%$ (solid blue) and $1\%$ (solid green) Rabi frequency crosstalk ratio. The error is given as a function of the difference in the compensation pulse phase from the optimal value, $\delta \phi = \Delta \phi - \pi$. The dashed lines indicate the results of a compensation pulse that is miscalibrated by $\pm20\%$ in amplitude. The shaded blue region represents the regime in which the compensation pulse will reduce the level of crosstalk.

We can now define the \textit{relative} crosstalk rotation error per $\pi$ pulse by taking the ratio of the compensated error (Eq. \ref{eq: pi pulse error}) to the original uncompensated crosstalk error (setting $f_{\rm comp} = 0$). If we assume that the level of crosstalk present in the system is weak, such that $f_{\rm CT} << 1$, then we can approximate this relative error simply as
\begin{equation} \label{eq: pi pulse error approx}
    \frac{\epsilon_{\pi}}{\epsilon_{\pi, 0}} \approx 1 + f_{\rm comp}^2 + 2 f_{\rm comp} \cos(\Delta\phi)
\end{equation}

This allows us to gain some intuition behind the robustness to miscalibrations of the phase and amplitude of the compensation pulse. If the compensation pulse amplitude is perfectly matched, the crosstalk error can be reduced by more than an order of magnitude ($\frac{\epsilon_{\pi}}{\epsilon_{\pi, 0}}$ = 0.1) for compensation phases that are miscalibrated by $\lesssim \pm 0.31$ rad, and likewise two orders of magnitude ($\frac{\epsilon_{\pi}}{\epsilon_{\pi, 0}}$ = 0.01) for phases $\lesssim \pm 0.09$ rad from the optimal value ($\Delta \phi = \pi$). If instead we assume that the compensation phase is at the ideal value, then for a relative crosstalk error suppression of one (two) order(s) of magnitude, the compensation pulse amplitude only needs to be calibrated to within $\lesssim \pm 32 \%$ ($\lesssim \pm 10 \%$).

\begin{figure}
\centering
    \includegraphics[scale=0.38]{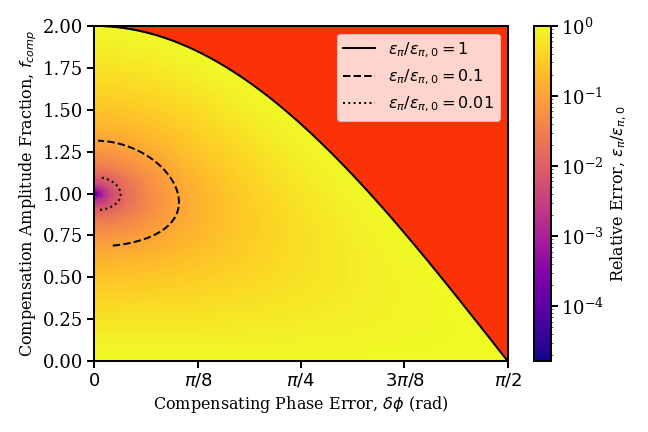}
    \caption{The relative rotation crosstalk error on the spectator ion per addressing $\pi$ pulse using PCC with a phase error from the optimal value, $\delta \phi = |\pi - \Delta\phi|$. The solid black line separates the break-even region of crosstalk error suppression, in which the red region represents an increase in the crosstalk error due to constructive interference with the compensation pulse. The dashed and dotted lines represent the threshold points where the relative error of crosstalk is reduced by a factor $10^1$ and $10^2$, respectively.}
    \label{fig: relative error}
\end{figure}

In Fig. \ref{fig: relative error} we plot the relative error per $\pi$ pulse (Eq. \ref{eq: pi pulse error approx}) as a function of the compensation phase and fractional compensation amplitude. The red region represents where the compensation pulse will increase the rotation error rather than reducing it. The dashed (dotted) line represents the region in which the crosstalk error is suppressed by a factor $10^1$ ($10^2$). We find a relatively large robustness to miscalibrations of both the amplitude and phase of the compensation pulse while maintaining large crosstalk suppression.  As we experimentally demonstrate in Section \ref{Sec: noise} and \ref{Sec: results}, the compensation phase and amplitude can be calibrated consistently well within the region of $>10^{3}$ relative crosstalk error suppression.

One last consideration is the crosstalk experienced by the target ion when applying a compensation pulse on the spectator ion. In our system, the relative Rabi ratio crosstalk is found to be $f_{\rm CT}=\Omega_{\rm CT}/\Omega_A \approx 0.1$ (see Section \ref{Sec: Exp setup} and \ref{Sec: results}). This implies that the back-action crosstalk experienced by the target ion due to the cancellation light should be expected to be on the order of $\sim10^{-4}$ Rabi ratio, and thus we considered it to cause negligible additional error on the target ion and do not further consider its effects.

\section{Experimental implementation} \label{Sec: noise}

\begin{figure}
\centering
     \begin{subfigure}[b]{0.4\textwidth}
         \centering
        \caption{}
        \includegraphics[scale=0.09]{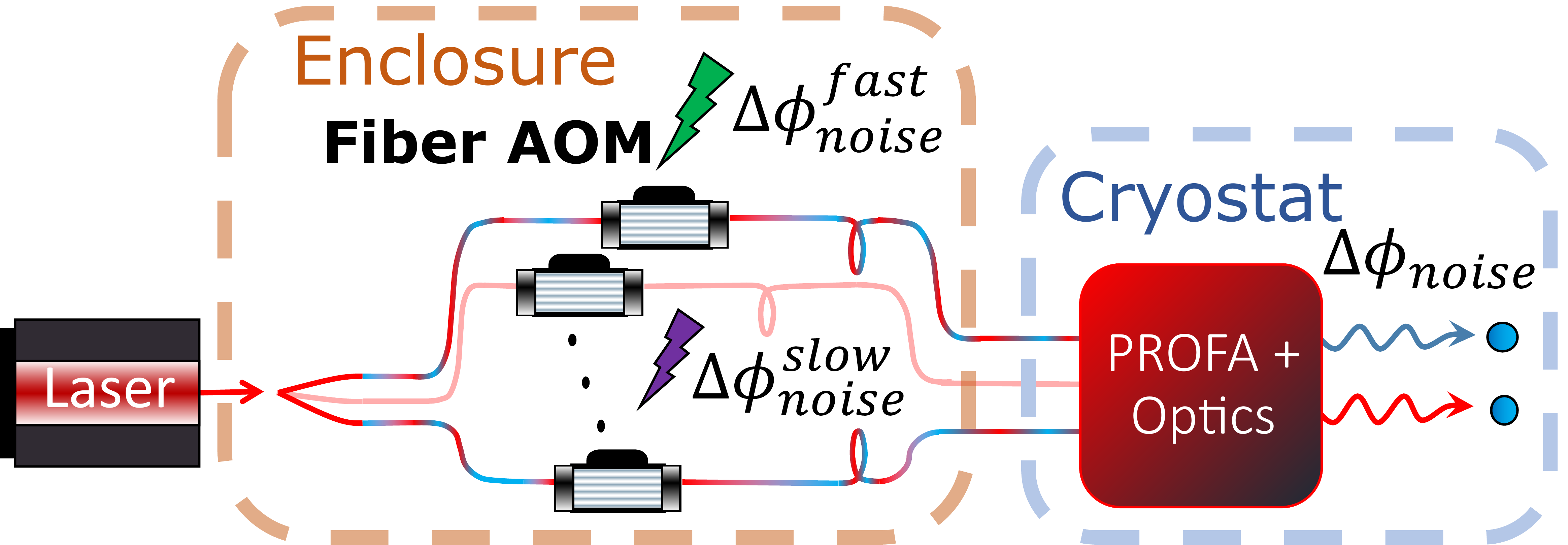}
        \label{fig: optic fiber setup}
     \end{subfigure} 
    \hfill
     \begin{subfigure}[a]{0.4\textwidth}
        \centering
        \caption{} %
        \includegraphics[scale=0.1]{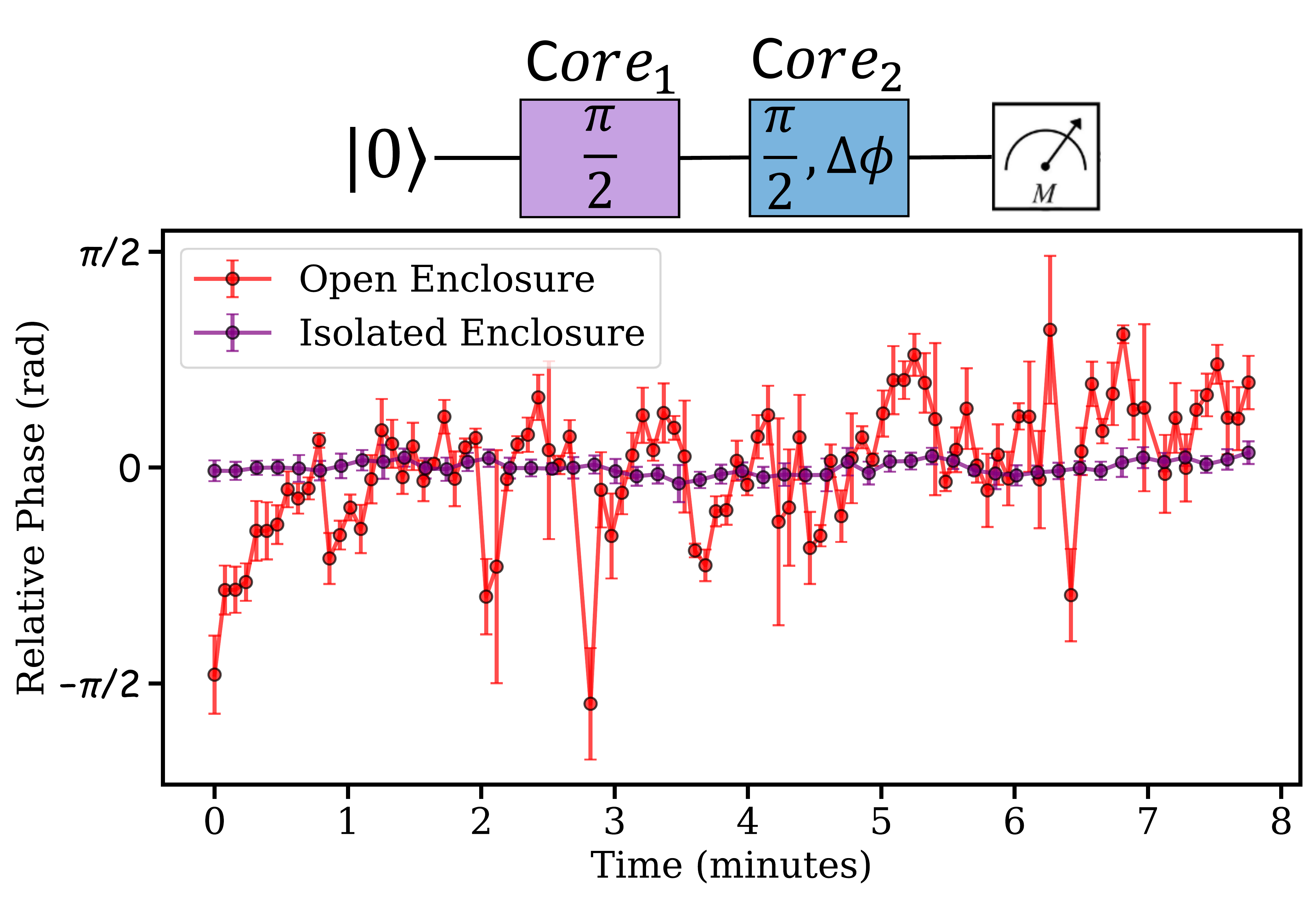}
        \label{fig: slow phase noise}
     \end{subfigure}
\caption{(a) Fiber optic setup of the addressing system. Fiber sections contained in the cryostat chamber (blue dashed box) are rigidly clamped to ensure that mechanical and thermal noise is common mode vibration noise, in order to attempt to impart negligible differential phase noise between fibers. Fiber sections exposed to the external ambient conditions (brown dashed box) exhibit both slow and fast noise caused by temperature/pressure fluctuations and duty cycle effects, respectively. (b) The relative phase between two addressing beam paths is probed using a Ramsey sequence with each Ramsey $\pi/2$ pulse from two different cores. The Ramsey fringe phase at zero wait time is repeatedly measured over time. This differential slow phase noise is plotted with (purple) and without (red) using an enclosure to isolate sections of fiber otherwise exposed to ambient conditions, resulting in a standard deviation in differential phase of $\sigma = 0.05$ rad and $\sigma = 0.49$ rad, respectively.}
\end{figure}

A sketch of the experimental setup for characterizing crosstalk cancellation is shown in figure Fig. \ref{fig: optic fiber setup}. The target channel and cancellation channel are fed from two different fibres, which are coupled into different channels of the PROFA device. Each fiber channel has an individual fiber acousto-optical modulator (AOM). Radio-frequency control applied to these fiber AOMs allow the power and phase of the light in both channels to be controlled precisely for each light path. Since the ion trap and PROFA device are mounted rigidly to the vacuum chamber inside the cryostat, all vibrations are common mode, while the optical path length between the ions and the PROFA outputs are phase coherent. However, the optical path length fluctuations due to different optical fibers may result in differential phase noise between optical channels.

We separate the optical fiber path into two regions. One section is the fiber bundle that enters the vacuum feedthrough propagating down to the 4 Kelvin stage within the cryogenic vacuum system (depicted within the blue dashed box of Fig. \ref{fig: optic fiber setup}). We find that the relative phase stability of this fiber section is minimal, which we think is due to the common path of all fibres in the bundle, and the limited capacity for differential thermal and vibration effects. The other section of fiber, which we find to be more susceptible to differential phase noise, is located in an isolated enclosure outside of the vacuum system. This enclosure includes the fiber splitters, polarization controllers, and separate fiber AOMs used to control the light pulses on each individual addressing beam (these are depicted within the brown dashed box in Fig. \ref{fig: optic fiber setup}).

We first consider passive drifts due to fluctuations of the temperature and air pressure under ambient conditions. We quantify these effects by monitoring the differential phase noise between two cores of the PROFA addressing device using a Ramsay experiment in which one core is used to perform the first $\pi/2$ pulse, and the other core to perform the second $\pi/2$. By monitoring the resulting spin population at zero Ramsey wait time, which depends on the relative phase as $P(\uparrow) = \frac{1}{2} ( 1 - \cos(\Delta \phi))$, we deduce the relative phases between the pulses from the two cores. This measurement is then performed over an extended period of time to monitor the long time drifts.

Results for our original configuration, with the AOMs laid out on an optical table are shown in the red data of Fig. \ref{fig: slow phase noise}. We find a standard deviation of $\sim0.49$ rad in differential phase fluctuations over a period of 8 minutes. We then used a metal housing to protect the relevant elements from ambient fluctuations. This enclosure was not hermetically sealed, but does include a lining of leaded foam, which aids to isolate the fiber from large air currents and sudden pressure changes as well as damping acoustic vibrations. We find this passive isolation significantly reduces the standard deviation of the differential phase measured over 8 minutes by a factor of $\sim10\times$ to a level of around $\sim 0.05$ rad (purple data in Fig. \ref{fig: slow phase noise}). It was found to drastically diminished the underlying slow drift in relative phase between cores to the order of $\sim 3.5\times10^{-3}$ rad/min. Such a drift rate would allow a factor of 100 reduction of the crosstalk error when using PCC, so long as the phase was re-calibrated every 25 minutes. 

In addition to these slow dynamics, we also observed fast dynamics related to pulsing the AOMs, which produced phase drifts that depend on the length and amplitude of the sequence of pulses used. We suspect that the RF drive to the fiber AOMs causes thermal effects that alter the optical path length of the light, resulting in a noticeable phase shift. In implementing quantum information operations for trapped ions, the on/off duty cycles of these AOMs can change considerably depending on the nature and type of sequences, such as the presence of sideband cooling pulses, the circuit depth of algorithmic sequences, and long idling times,  for example in probes of coherence using Ramsey experiments.

\begin{figure}
\centering
    \includegraphics[scale=0.2]{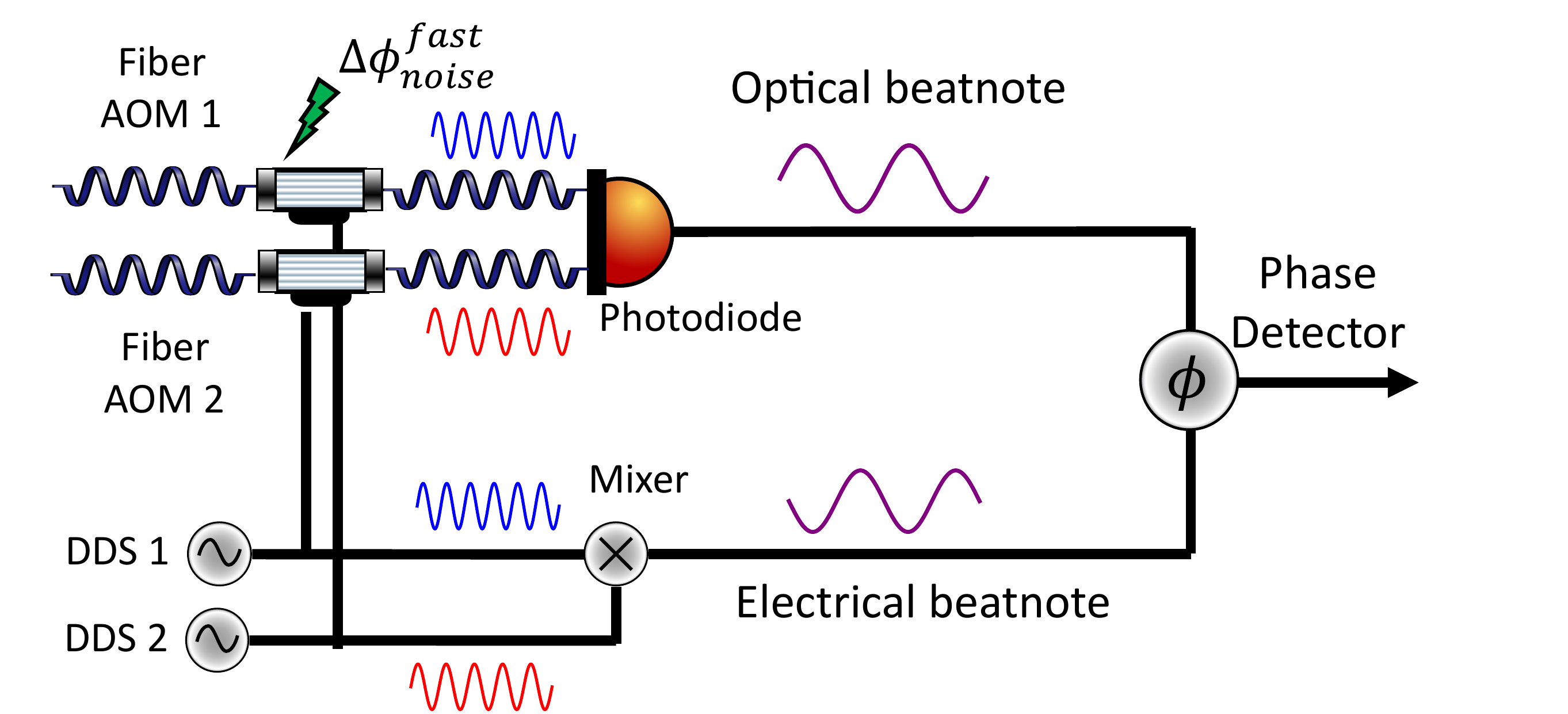}
    \caption{Diagram depicting the setup used for the measurement of fast duty cycle-related phase noise. The two DDS channels used to drive the two core fiber AOMs are mixed to produce an electrical beatnote at 2 MHz. The light from the two AOMs are optically combined onto a photodiode to create an optical beatnote. The phases of the electrical and optical beatnotes are then compared with a phase detector.}
    \label{fig: duty cycle phase effects}
\end{figure}

To study these effects, we used the setup shown in Fig. \ref{fig: duty cycle phase effects}. Two separate direct digital synthesizers (DDS) were used to produce RF signals that drive two individual fiber AOMs. The two AOMs are driven at slightly different frequencies, $\Delta f$, and are electronically mixed together to create a reference beatnote of $\cos(2\pi\Delta f t)$. The two optical outputs of the fiber AOMs are also optically mixed on a photodiode to create an optical signal at the same frequency as the reference. The optical phase noise, $\Delta\phi_{noise}$, that is acquired from the different light paths of the two cores will then be imprinted onto this optical beatnote resulting in a signal proportional to $\cos(2\pi\Delta f t + \Delta\phi_{noise})$. We then compare these two signals electrically using a phase detector to measure this differential phase noise between cores.

\begin{figure}
\centering
    \includegraphics[scale=0.14]{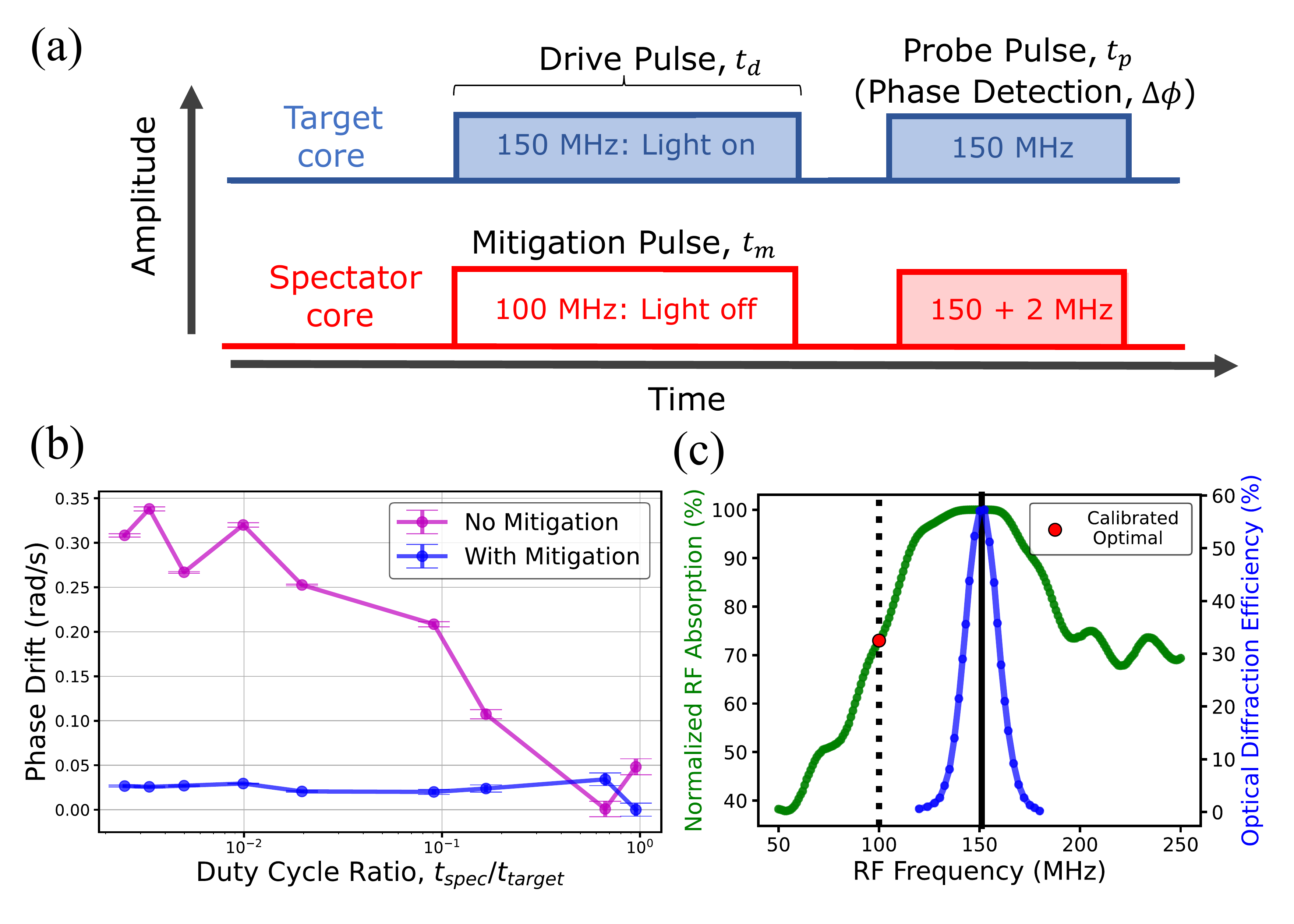}
    \caption{(a) Pulse sequence used to measure differential phase noise between the target core (blue) and spectator core (red) for a varying drive pulse length, $t_d$, to alter the duty cycle ratio, followed by a probe pulse to detect the differential phase at a fixed pulse length $t_p =$100 $\mu$s. This sets the spectator optical pulse time to be $t_{spec} = t_p$, while the variable target pulse length is $t_{target} = t_p + t_d$. The shaded region shows when the fiber AOMs are driven on or near resonance in which light is still efficiently diffracted through the fiber. The non-shaded mitigation pulse on the spectator core is applied far off-resonance at 100 MHz where negligible amounts of light are transmitted through the fiber AOM. (b) Phase drift rate observed while operating with a duty cycle ratio between the times of the spectator, $t_{spec}$, and target core, $t_{target}$, optical pulses with (blue) and without (purple) applying the fast duty cycle phase noise mitigation technique as described in the text. (c) Measured optical diffraction efficiency (blue) and RF absorption (green) as function of the RF drive frequency. The optimal power to drive the resonant pulse is found by the relative power absorption at the far detuned frequency (black dotted line). 
    }
    \label{fig: mitigation method}
\end{figure}

To initially characterize the system, we start by applying a fixed Radio Frequency (RF) amplitude ``drive'' pulse with a variable duration, $t_d$, solely to the target ion core at the resonant frequency of the fiber AOM, set at 150 MHz. This pulse sets the relative duty cycle ratio. Following this, the differential phase is measured with a brief 100 $\mu$s "probe" pulse, $t_p$, administered to both the target and spectator cores, operating at 150 MHz and 152 MHz, respectively, creating electrical and optical beatnotes at a frequency difference, of $\Delta f =2$ MHz. This is illustrated by the shaded blue and red pulse sequence in Figure \ref{fig: mitigation method}a. The difference in optical phase between the cores is then captured by comparing the phases of these electrical and optical beatnotes. This procedure is repeated for many shots, revealing an initial slow linear drift in the measured differential phase on the order of 10 seconds before stabilizing.

This initial linear phase drift rate is indicated by the purple data in Figure \ref{fig: mitigation method}b. We define the duty cycle ratio as the ratio between the optical pulse lengths applied to spectator and target cores, $t_{spec}/t_{target}$. Since the phase probe pulse is uniformly applied to both cores at $t_p=$ 100 $\mu$s, and the drive pulse is solely administered to the target ion to manipulate the duty cycle ratio, we have $t_{spec} = t_p$ and $t_{target} = t_d + t_p$. By altering the duration of the drive pulse between 5 $\mu$s and 40 ms, we can adjust the duty cycle ratio. As the duty cycle ratio approaches unity, we observe a decrease in the relative phase drift rate, aligning with expectations.

To address the duty cycle effects, we employ a strategy to constantly load the fiber AOMs with a high duty cycle, regardless of whether they are actively diffracting light. This approach aims to minimize relative temperature fluctuations in pulsed mode \cite{Wang2022}. We achieve this by applying Radio Frequency (RF) power to the fiber AOM controlling the spectator ion core in parallel with the pulse in the target ion fiber AOM. This is set to put the same thermal load on the spectator AOM as in the target while not diffracting light in the former. The parameters of the two tones are selected based on considerations outlined in Figure \ref{fig: mitigation method}c. The non-diffractive RF pulses applied to the spectator at 100 MHz (indicated by the black dotted line) fall within the RF bandwidth (as depicted by the absorption spectrum in green) of the AOM driving circuitry but are detuned compared to the optical diffraction bandwidth of the AOM (shown in blue), resulting in minimal light transmission. This allows the RF duty cycle mitigation pulse to heat the AOM while preventing the unwanted application of gates. For our calibrations, the target core pulse at 150 MHz (shown by the black solid line), where light is diffracted and the RF power is highly absorptive, is adjusted to match the absorbed power of the spectator at the far-detuned pulse where the RF absorption is lower, which is applied at full power. In these experients we used the off-resonant tone to only mitigate duty cycle effects, however. A further upgrade would be to use  two RF tones: one weak tone near resonance for applying crosstalk cancellation, and one strong RF tone off-resonance that acts to match the heat load of the target ion AOM, thereby reducing duty cycle effects.

With this mitigation approach, we observe a reduction in the phase drift rate by an order of magnitude to approximately 0.035 rad/s, as depicted in Figure \ref{fig: mitigation method}b (blue data), a level
at which calibrations can maintain significant crosstalk
suppression. These results were achieved using the maximum RF power to optimize optical power efficiency. Reducing the operational pulse powers will also decrease the resulting duty cycle heating effects.

\section{Characterization of Crosstalk Errors} \label{Sec: results}

In order to investigate the effectiveness of the optical PCC technique, we compare its performance to other techniques. Specifically, composite pulses have been employed to reduce the detrimental effects of crosstalk \cite{Egan2021, Merrill2014}, and there are several proposed sequences for both single qubit \cite{Ivanov2011} and two qubit gates \cite{Torosov2023} that can suppress addressing crosstalk. These composite pulse sequences, which we will refer to as \emph{algorithmic} techniques for crosstalk suppression, can act as a narrowband or bandpass filter in which erroneous rotation errors at low driving amplitudes are suppressed relative to the full area single-qubit gate pulse.

\begin{figure}
\centering
     \begin{subfigure}[a]{0.4\textwidth}
        \centering
        \caption{} \label{fig: algorithmic vs physical}
        \includegraphics[scale=0.1]{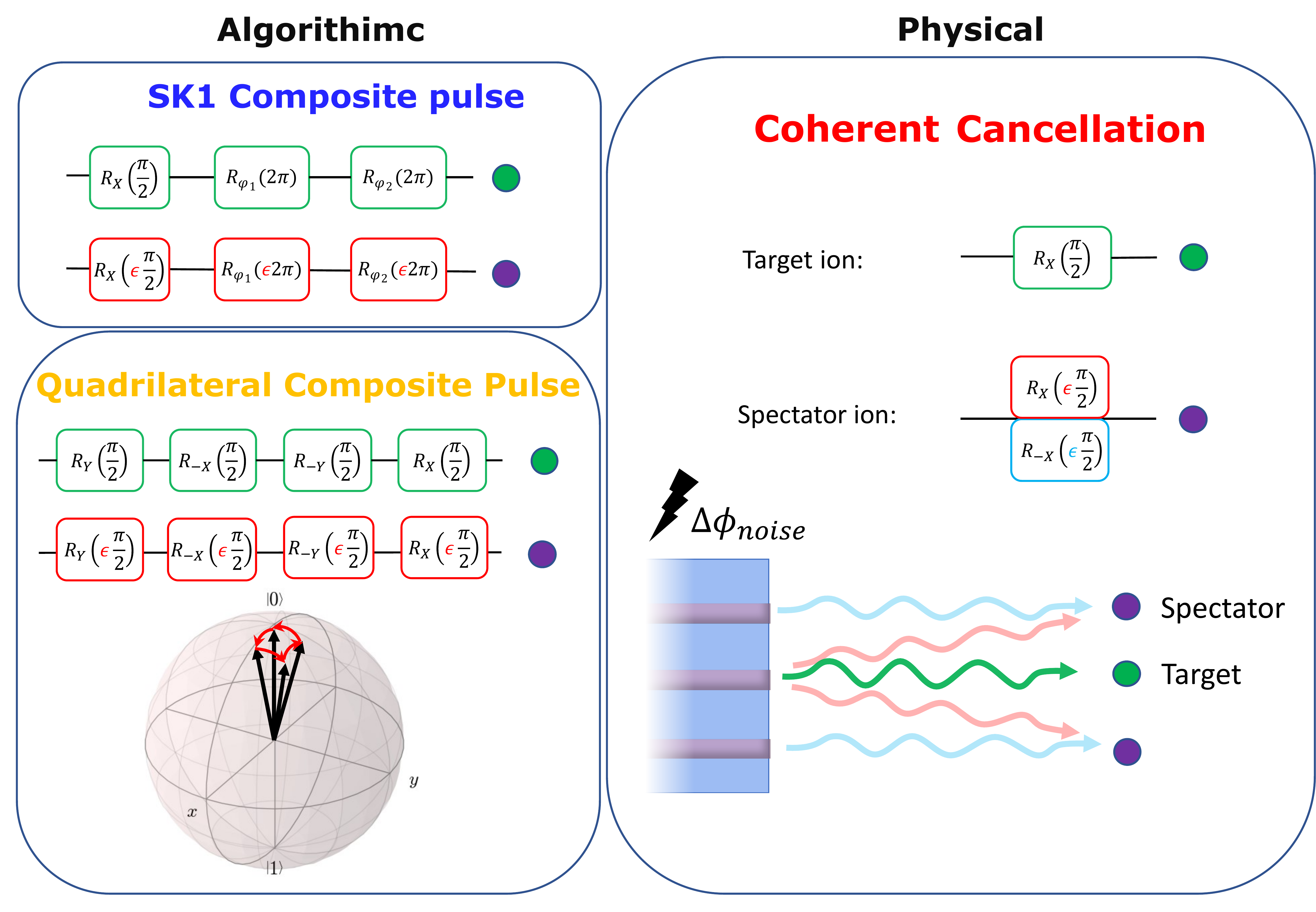}
     \end{subfigure}
    \hfill
     \begin{subfigure}[b]{0.4\textwidth}
         \centering
        \caption{} \label{fig: composite pulse intensity}
        \includegraphics[scale=0.4]{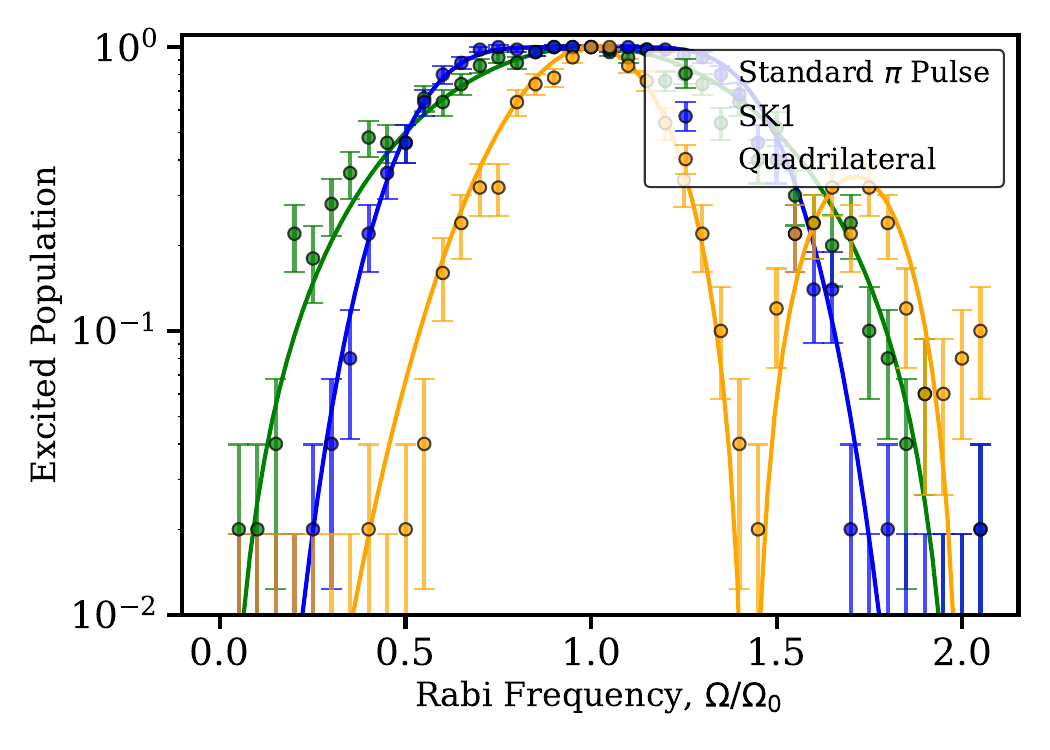}
     \end{subfigure}
     \caption{(a) The two types of crosstalk mitigation techniques compared in this work: \textit{Algorithmic} and \textit{Physical} compensation. The SK1 and quadrilateral composite pulse sequences are comprised of rotations, $R_\phi(\theta)$, with angle $\theta$ about an axis define by $\phi$ on the target ion (green). Crosstalk on the spectator ion is described by weak rotations (red) in which $\epsilon = \Omega_{\rm CT}/\Omega_0 \ll 1$. The action of the quadrilateral sequence on a spectator ion can be viewed on the Bloch sphere to undergo a small square trajectory, returning approximately to its initial state. The PCC method instead uses a second weak pulse applied to the spectator ion to cancel crosstalk intensity. (b) Excited state population after a $\pi$ pulse implemented using a standard square pulse of length $t_{\pi}$ as well as the SK1 and quadrilateral composite pulses for varying relative Rabi frequencies such that $\Omega_0 = \pi/t_{\pi}$. The solid lines are theory calculations. The asymmetry of the quadrilateral pulse is due to the off-resonant drive at large powers as the qubit frequency is Stark shifted.}
\end{figure}

We employed two composite pulse sequences, which are illustrated in Fig. \ref{fig: algorithmic vs physical}. The first is the SK1  \cite{Brown2004}, which has the benefit of being robust against amplitude noise on the target ion qubit gates as well as acting to suppress crosstalk errors on neighbouring ion sites experiencing a weak drive.
These qualities make the SK1 sequence very appealing and have led to their being used for schemes requiring high fidelity operations in longer ion strings for quantum computing protocols \cite{Egan2021, Chen2023}. 

In an attempt to improve the performance of algorithmic techniques we also introduce a custom sequence we will refer to as the \textit{quadrilateral} composite pulse sequence. This pulse sequence is described by four segments of unitary rotations with angle, $\theta$, about the X and Y axis in the Bloch sphere, $\hat{R}_{\pm X, Y}(\theta)$, as
\begin{equation} \label{eq: Quad}
    \hat{U}_{Quad}(\epsilon) = \hat{R}_Y\left(\epsilon \frac{\pi}{2}\right) \hat{R}_{-X}\left(\epsilon \frac{\pi}{2}\right) \hat{R}_{-Y}\left(\epsilon \frac{\pi}{2}\right)\hat{R}_X\left(\epsilon \frac{\pi}{2}\right)
\end{equation}
for a relative Rabi frequency of $\epsilon = \frac{\Omega}{\Omega_0}$ with respect to an ideal $\pi/2$ pulse Rabi frequency $\Omega_0$. For the target qubit, $\epsilon \approx 1$ and thus $\hat{U}^{Target}_{Quad} \approx \hat{R}_X(\pi/2)$. The spectator ion will instead undergo a drive with $\epsilon = \frac{\Omega_{\rm CT}}{\Omega_0} \ll 1$, in which each segment can be approximated by the rotation $\hat{R}_{\hat{P}} \approx \hat{I} + i(\frac{\pi}{4} \epsilon)^2\hat{P}$ for the Pauli operators $\hat{P} = X, Y$. To highest order, Eq. \ref{eq: Quad} then reduces to $\hat{U}^{Spectator}_{Quad} \approx \hat{I} + \mathcal{O}(\epsilon^4)$ for the spectator ion dynamics. This sequence serves as a simple higher order narrow-band sequence, although there are many other types of higher-order sequences that have also been explored \cite{Merrill2014}.

The trajectory of the spectator ion under the effect of the quadrilateral pulse sequence can be easily visualized on the Bloch sphere under the small rotation angle approximation as a simple square about its initial state, as shown in the inset of Fig. \ref{fig: algorithmic vs physical} for the initial $|0\rangle$ state. Although the crosstalk suppression is higher order than for SK1, this sequence does not provide robustness with regards to intensity fluctuations on the targeted ion. Consequently, the quadrilateral sequence may be less useful for implementation in actual circuits. Fig. \ref{fig: composite pulse intensity} shows experimental as well as theory results comparing the sensitivity to Rabi frequency fluctuations of a standard square $\pi$-pulse (green) vs the  SK1 (blue) and quadrilateral composite pulses (yellow). Two parameter regimes are of interest. The first is close to a Rabi frequency producing a $\pi$ pulse, where we observe the insensitivity of SK1 and increased sensitivity of the quadrilateral pulse. The other regime is where the Rabi frequency  is much less than the target $\Omega/\Omega_0 \ll 1$, which is the typical regime of crosstalk. In this regime we see the the composite pulse sequences effectively reduce the crosstalk.

We calibrate the PCC pulses by first determining the duration of pulse required to perform a $\pi$ pulse on the spectator ion, $t^{\rm CT}_{\pi}$ due to crosstalk. We then calibrate the power of the cancellation pulse to match the measured $\pi$ time. Finally, the phase of the compensation pulse is calibrated by applying both the target ion gate and spectator cancellation pulses simultaneously. The phase of the cancellation pulse can then be scanned and the resulting data fitted using the theoretical prediction of Eq. \ref{eq: error} to extract the correct compensation phase (as previously shown in Fig. \ref{fig: phase calibration} by the dotted black line at a pulse time of $t = 2t^{\rm CT}_{\pi}$). This calibration can be made more accurate by using longer pulse times.

\begin{figure*}[ht]
\centering
     \begin{subfigure}[a]{0.45\textwidth}
        \centering
        \caption{} \label{fig: x errors}
        \includegraphics[scale=0.16]{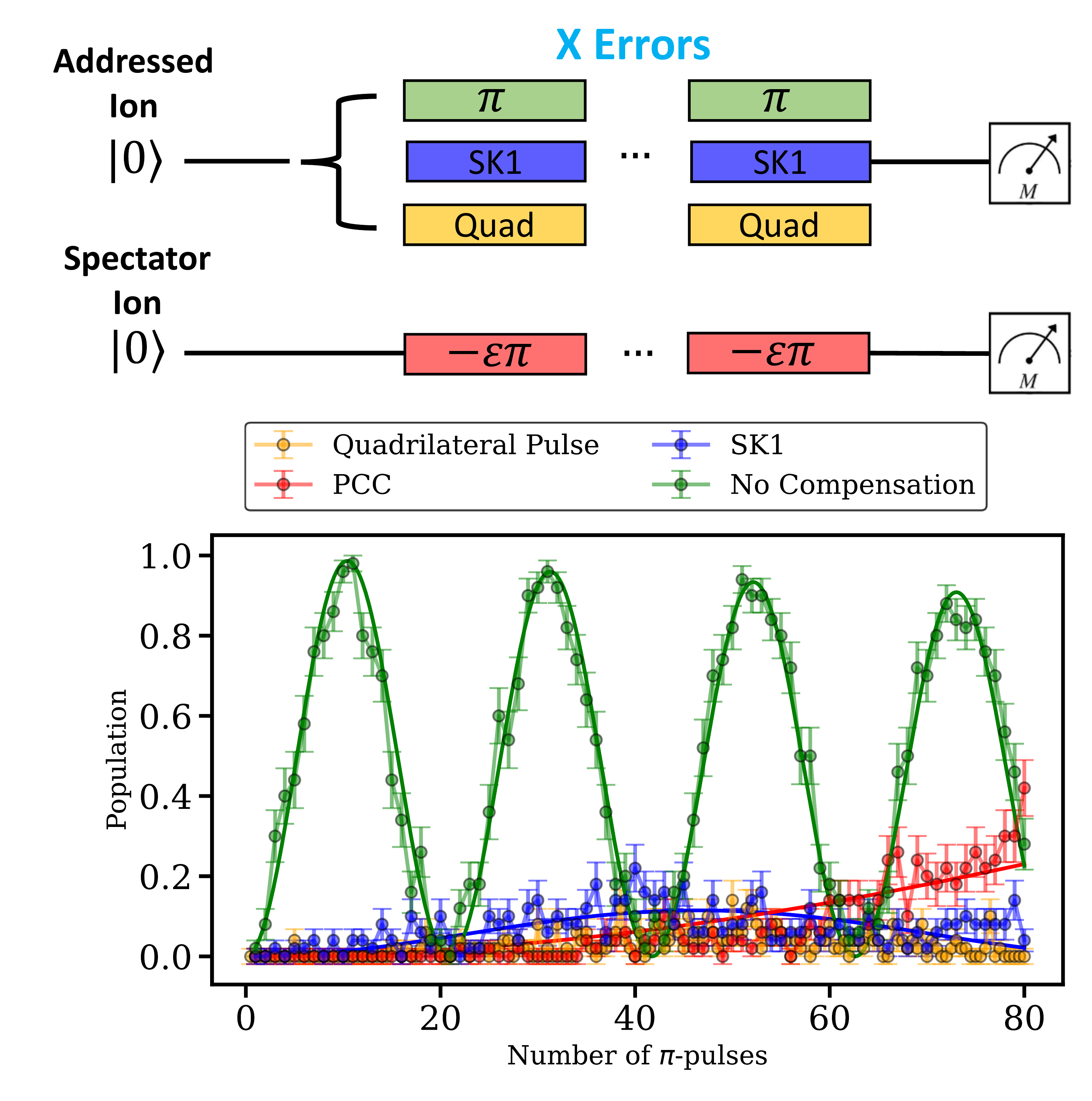}
     \end{subfigure}
    \hfill
     \begin{subfigure}[a]{0.5\textwidth}
         \centering
         \caption{} \label{fig: z errors}
        \includegraphics[scale=0.16]{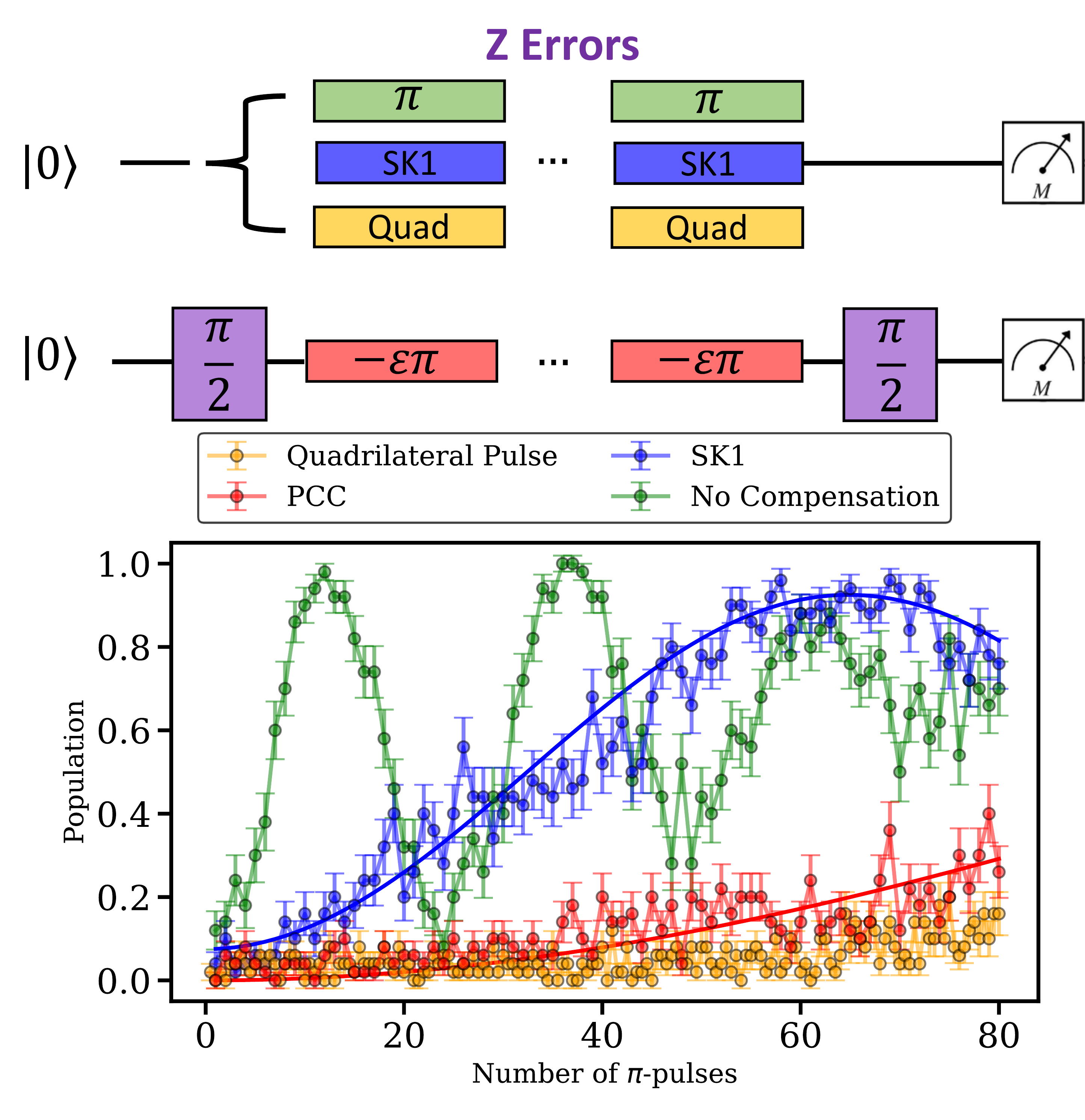}
     \end{subfigure}
\caption{The benchmarking pulse sequence for (a) $X$-type errors and (b) $Z$-type errors are shown at the top of each plot. The excited state populations of the spectator ions are measured after a sequential application of $N$ target ion $\pi$ pulses while the spectator ion is prepared in the (a) $Z$ and (b) $X$ eigenstates. The solid lines are theoretical fits of the given pulse sequence, using fitting parameters of effective crosstalk ratio, $f_{eff}$, and detuning, $\Delta_{CT}$. Both without the any crosstalk compensation (green) as well as the PCC (red) method were fit with the simple error model Eq. \ref{eq: pi pulse error}, while the SK1 (blue) method is fit numerically to the composite pulse sequence. Both the rotation ($X$-type) and phase ($Z$-type) bare crosstalk errors on the spectator ion (shown in green) are suppressed using the PCC technique (red) as well as for the SK1 (blue) and quadrilateral (yellow) composite pulse schemes.}
\end{figure*}

To characterize the errors caused by crosstalk we measure both $X$-type errors, caused by resonant rotations, and $Z$-type errors, resulting from qubit phase shifts due to AC Stark shifts. Since the cross-talk fields are generally small, the latter are primarily due to the detuning between the target ion transition (which is itself AC Stark shifted) and the un-shifted spectator qubit frequency. $X$-type errors are determined by measuring the excited state population of the spectator ion after applying many $\pi$-pulses on the target ion. The $Z$-type errors are instead determined by performing a Ramsey experiment on the spectator ion in which a $\pi/2$ pulse is applied to the spectator ion before and after the train of target ion $\pi$-pulses are applied. This acts to measure the spectator ion in the $X$ basis in the presence of off-resonant crosstalk light.
Fig. \ref{fig: x errors} and \ref{fig: z errors} show these two experimental procedures for measuring the $X$ and $Z$ type errors, respectively, as well as experimental results comparing PCC and algorithmic methods. The PCC technique is compared to the two types of algorithmic techniques previously described. The quadrilateral pulse only produces a $\pi/2$ rotation, so we apply it twice to realize a corresponding $\pi$ pulse.

The uncompensated Rabi oscillations ($f_{comp}=0$) that the spectator ion undergoes due to crosstalk is shown in green in Fig. \ref{fig: x errors}. By fitting the simple model of Eq. \ref{eq: pi pulse error}, we find that the crosstalk ratio is $f_{CT}=\Omega_{\rm CT}/\Omega_0\approx0.096$, which produces an $X$ error of $\epsilon_{\pi}\approx 2.3\times10^{-2}$ per $\pi$ pulse. When PCC is applied, we have observed that the remaining effective crosstalk Rabi ratio, $f_{eff}$, can be reduced by a factor of $\sim25-40$ corresponding to the crosstalk light intensity being suppressed on the order of $\sim10^3$. Again using Eq. \ref{eq: pi pulse error} to fit the PCC method (Fig. \ref{fig: x errors}, red data) we find that $f_{eff} = \Omega_{eff}/\Omega_0 \approx 4\times 10^{-3}$, corresponding to a rotation error per $\pi$ pulse of only $\epsilon^{PCC}_{\pi} \approx 3.9\times10^{-5}$, providing more than two orders of magnitude in crosstalk error suppression. The Quadrilateral composite pulse sequence (yellow data) is found to produce a similar amount rotation error. Conversely, the SK1 composite pulse is performed (blue data), and numerically simulated (solid blue line) by fitting for a relative Rabi frequency and detuning. We find the SK1 sequence instead gives effective error per target ion $\pi$ pulse of $\epsilon^{SK1}_{\pi}\sim 1.3 \times 10^{-4}$.

The comparison of PCC to the composite pulses for the off-resonant $Z$-type errors shows a much stronger variation in behaviour, as shown in Fig. \ref{fig: z errors}, and requires a more subtle interpretation. This experiment prepares and measures the state in the X-basis, however, since the crosstalk pulses are still a near-resonant drive, they still provide information about the rotational errors as well as the phase errors. The green data, displaying the bare spectator population without any crosstalk compensation, is found to produce oscillations similar in strength to that found in Fig. \ref{fig: x errors}, suggesting that the X rotations are the dominate source of error.

For the PCC method (Fig. \ref{fig: z errors}, red data) we find a similar effective crosstalk as before, with $f_{eff} = 4.5\times 10^{-3}$ giving an error per $\pi$ pulse of $\epsilon^{PCC}_{\pi}\approx5.1\times10^{-5}$. This implies that the residual error is still mostly due to $X$ rotations. The SK1 sequence (blue) does not exhibit robustness to crosstalk errors in this basis, as it is designed assuming an equal target and spectator ion qubit frequency. Simulating these dynamics again we now find that the error per $\pi$ pulse is significant at $\epsilon^{SK1}_{\pi}\approx2.6\times10^{-2}$. The quadrilateral sequence (yellow) is shown to be effective against errors in this basis as well, as long as the correct AC Stark shift of the target ion is calibrated and the appropriate phase in the Pauli frame update is used. PCC thus provides a means to mitigate crosstalk rotation error at least as effectively as composite pulses, without the need for additional circuit depth overhead.

One subtlety of the PCC method is that a mismatch in the polarization between the target crosstalk and cancellation light will also limit the PCC compensation fidelity. To mitigate this, polarization is initially calibrated by matching transition strengths on a neighbouring ion pair using an in-line fiber optic polarization controller. The results presented in this paper were taken with the polarization only calibrated once and may be subject to drift, particularly over long time periods.

\section{Conclusion}

We have characterized physical crosstalk compensation by applying a coherent pulse of light on the spectator site to destructively interfere cross illumination from a target pulse. We find this method of optical crosstalk cancellation can allow for crosstalk intensity to be suppressed by a factor $>10^3$, with a reduction in the $X$ rotation crosstalk error by a factor of $>500$. This is comparable to the levels we achieve through algorithmic composite-pulse techniques such as SK1, however, we find that the physical cancellation of crosstalk produces much lower $Z$ phase errors than this leading alternative pulse sequence.

Sources of error of the PCC technique due to differential phase noise between the target and compensation pulses were also investigated. We identified and mitigated noise from both slow drifts caused by ambient fluctuations of temperature and pressure, as well as fast phase noise from duty cycle effects. We find that applying an off-resonant tone to the driving AOMs result in a steady state duty cycle which can produce a $>10$ times reduction in phase drift rates, allowing for a routine calibration of cancellation pulses to be feasible.

Since this scheme only relies on the coherence between the crosstalk and the compensation pulse, it can provide a general solution to crosstalk mitigation that may be employed by various quantum computing platforms in addition to trapped-ions. These findings may also be relevant for other ion trap setups that require phase-stable operations such as free-space single ion addressing and integrated optical waveguide surface traps \cite{Mehta2016, Mehta2020, Ivory2021, Binai2023}. For systems with only nearest neighbour crosstalk issues this scheme may provide a scalable means of negating errors due to crosstalk which may otherwise compromise the fault tolerance of a quantum computing circuit.

\section{Acknowledgments}\label{Sec:Acknowledgements}
The authors would like to thank \textit{Chiral Photonics, Inc.} for providing the multi-core photonic crystal fiber used in the single ion addressing system. 

We acknowledge funding from the Swiss National Fund under Grant numbers $200020\_207334$ and $200020\_179147$, and from the EU Quantum Flagship H2020-FETFLAG-2018-03 under Grant Agreement No. 820495 AQTION, as well as from Intelligence Advanced Research Projects Activity (IARPA), via the US Army Research Office grant W911NF-16-1-0070.

This work was supported as a part of NCCR QSIT, a National Centre of Competence (or Excellence) in Research, funded by the Swiss National Science Foundation (grant number 51NF40-185902).

\bibliography{BibFile}

\end{document}